\documentclass[aps,prc,preprint,showpacs,groupedaddress,preprintnumbers,
amsmath,amssymb]{revtex4}
 
\usepackage{graphicx}
\usepackage{bm}

\begin{document}

\title{ $\alpha$-particle photoabsorption with a realistic nuclear force}

\author{Doron Gazit$^{1}$, Sonia Bacca$^{2}$, Nir Barnea$^1$, \\
Winfried Leidemann$^{3}$, 
and Giuseppina Orlandini$^{3}$}

\affiliation{$^1$The Racah Institute of Physics, The Hebrew University, 
     91904 Jerusalem, Israel}

\affiliation{$^2$Gesellschaft f\"ur Schwerionenforschung,
Planckstr.~1, 64291 Darmstadt, Germany
}

\affiliation{$^3$Dipartimento di Fisica, Universit\`a di Trento and Istituto Nazionale di Fisica Nucleare, Gruppo Collegato di Trento, 
I-38050 Povo, Italy}

\date{\today}

\begin{abstract}
The $^4$He total photoabsorption cross section is calculated with the realistic 
nucleon-nucleon potential Argonne V18 and the three-nucleon force (3NF)  
Urbana IX. Final state interaction is included rigorously via the Lorentz 
Integral Transform method. A rather pronounced giant resonance with peak cross sections 
of 3  (3.2) mb is obtained with (without) 3NF. Above 50 MeV strong 3NF effects, up to 
35\%, are present. Good agreement with experiment is found close to 
threshold. A comparison in the giant resonance region is inconclusive,
since present data do not show a unique picture.
\end{abstract}

\maketitle

In the last three decades there has been a continous interest in $^4$He 
photodisintegration, both in theory and in experiment (see  
\cite{Lund,Shima} and references therein). The $\alpha$-particle is
drawing such a great attention because it has some typical features of heavier 
systems (e.g. binding energy per nucleon), which make it an important link 
between the classical few-body systems, i.e. deuteron, triton  and $^3$He, and more 
complex nuclei. For example in $^4$He one can study the possible emergence of 
collective phenomena typical of complex nuclei like the giant resonance.  
Furthermore $^4$He is the 
ideal testing ground for microscopic two- and three-body forces,  which are 
fitted in the two- and three-body systems. At present the 3NF is not yet well determined,
thus it is essential to search for observables where it plays an important role.
Because of gauge invariance, in electromagnetic processes nuclear forces manifest 
themselves  also as exchange currents, which have turned out to be very important
in photonuclear reactions and hence 3NF effects might become significant. 
For the three-nucleon systems photonuclear processes have already been 
studied \cite{ELOT,Fad_CHH,RepBoKr}, e.g. in \cite{ELOT} it was found that the 
3NF leads to an almost 10\% reduction of the electric dipole peak 
and up to 15\% enhancement  at higher energy. One expects that the 3NF is  
of considerably greater relevance in the four-body system, since there one has six
nucleon pairs and four triplets, compared to three pairs and just one triplet
in the three-nucleon systems.

Unfortunately, the current theoretical and experimental situation of the $^4$He 
photodisintegration is not sufficiently settled. Most of the experimental work has 
concentrated on the two-body break-up channels $^4$He$(\gamma,n)^3$He and
$^4$He$(\gamma,p)^3$H in the giant resonance region, but still today there is large 
disagreement in the peak. In fact in two very recent $(\gamma,n)$ experiments 
\cite{Lund,Shima} one finds differences of a factor of two. 
The $^4$He$(\gamma$) reaction represents a very challenging theoretical problem as well, 
since the full four-body continuum 
dynamics and all possible fragmentations have to be considered. Calculations with 
realistic nuclear forces have not yet been carried out. Even investigations with  
a realistic nucleon-nucleon (NN) interaction and without a 3NF are still missing. 
Calculations exist only for semi-realistic NN potentials. In Refs. 
\cite{ELO97,BELO01,Sofia1} it has been shown that such models lead to pronounced 
peak cross sections, in rather good agreement with the experimental data of 
\cite{Lund} and much different from what was calculated earlier \cite{Sandhas}.

It is evident that the experimental and theoretical situations are very unsatisfactory. 
With the present work we make an important step forward on the theory side 
performing a calculation of the $\alpha$-particle total photoabsorption cross 
section with a realistic nuclear force. To this end we solve the four-body
problem taking as nuclear interaction the realistic Argonne V18 (AV18) NN potential 
\cite{AV18} and the Urbana IX (UIX) 3NF \cite{Urb9}. We reduce the continuum state 
problem to a bound-state like problem via the Lorentz Integral Transform (LIT) 
method \cite{ELO94}, thus taking into account the full final state interaction 
rigorously. The LIT bound-state like problem, as well as that of the $^4$He 
ground state, is solved via expansions in hyperspherical harmonics (HH) using the 
powerful effective interaction HH (EIHH) approach \cite{EIHH,EIHH_3NF}. The reliability 
of the EIHH technique in electromagnetic processes with the AV18 and UIX potentials 
has been shown in a recent $^3$H$(\gamma)$ calculation  \cite{BLOET05}.

We calculate the total photoabsorption cross section $\sigma_\gamma(\omega)$ in 
the unretarded dipole approximation. In this way the dominant part of the exchange 
current contribution on the total cross section is taken into account via the 
Siegert theorem. In the classical few-body systems it has been found 
\cite{Fad_CHH,SaA} that this is an excellent approximation, particularly for 
photon energies $\omega$ below 50 MeV. For example for the triton
$\sigma_\gamma(\omega)$  the contributions of retardation and all other multipoles 
lead to a cross section enhancement of less than 1\% for $\omega 
\le 40$ MeV and of 5 (5)\%, 16 (18)\%, and 26 (33)\% with AV18 (AV18+UIX) at 
$\omega= 60$, 100, and 140 MeV, respectively \cite{Fad_CHH}. 

The total photoabsorption cross section is given by 
\begin{equation}
\sigma_\gamma(\omega)=4\pi^{2}\alpha\omega R(\omega)\,,
\end{equation}
\noindent where $\alpha$ is the fine structure constant and
\begin{equation} \label{1}
   R(\omega )=\int d\Psi _{f}\left| \left\langle \Psi _{f}\right| \hat{D}_z \left| 
\Psi _{0}\right\rangle \right| ^{2}\delta (E_{f}-E_{0}-\omega) 
\end{equation}
is the response function in the unretarded dipole approximation with 
$\hat{D}_{z}=\sum_{i=1}^{A}\frac{\tau^{3}_{i}z'_{i}}{2}$. The wave functions of 
the ground and final states are denoted by $\left| \Psi_{0/f} \right>$ and the 
energies by $E_{0/f}$, respectively. The operators $\tau^{3}_{i}$  and $z'_{i}$
are the third components of the $i$-th nucleon isospin and center of mass frame 
position. In the LIT method  one obtains $R(\omega)$ from the inversion of an 
integral transform with a Lorentzian kernel \cite{ELO94}
\begin{equation}
L(\sigma_{R},\sigma_{I} )= \int d\omega \frac{R(\omega )}
{(\omega -\sigma _{R})^{2}+\sigma ^{2}_{I}}= \left\langle 
\widetilde{\Psi }|\widetilde{\Psi }\right\rangle \,, \label{2}
\end{equation}
\noindent where $\widetilde\Psi$ is the  
unique solution of an inhomogeneous ``Schr{\"o}dinger-like'' equation 
\begin{equation}
\label{3}
(H-E_{0}-\sigma_{R}+i\sigma_{I})|\widetilde{\Psi}\rangle=\hat{D}_{z}|
{\Psi_{0}}\rangle
\end{equation}
\noindent 
with bound-state like asymptotic boundary conditions. 

The EIHH expansions of $\Psi_{0}$ and $\widetilde {\Psi}$ are performed with
the full HH set up to maximal values of the HH grand-angular momentum quantum 
number $K$ ($K\leq K^0_{m}$ for $\Psi_0$, $K\leq K_{m}$ for $\widetilde\Psi$). 
The convergence of binding energy and matter radius are presented in Table I. 
Our final binding energy results of 24.27 MeV (AV18) and 28.42 MeV (AV18+URBIX) agree 
quite well with other calculations (AV18: 24.25 \cite{Fad}, 24.22 \cite{LaC}, 
24.21 \cite{Pisa} MeV); AV18+URBIX: 28.34 \cite{GFMC}, 28.50 \cite{Fad}, 28.46 
\cite{Pisa} MeV).

The EIHH convergence of the transform $L$ is excellent for the AV18 potential. 
Here we discuss in detail only the case AV18+UIX, where the convergence is quite 
good,  but not at such an excellent level. The reason is that in our 
present EIHH calculation an effective interaction is constructed only for the 
NN potential, while the 3NF is taken into account as bare interaction. In Fig.~1 
we show results for the transform $L$ obtained with various $K_m$ and $K^0_m$. 
Since we take $K^0_{m}=K_{m}-1$ the corresponding transform can be denoted by 
$L_{K_{m}}$. One sees that there is a very good convergence beyond the peak. 
The figure also shows that the peak height is very well established, but that 
the peak position is not yet completely converged. In fact with 
increasing $K_{m}$ the peak is slightly shifted towards lower $\sigma_R$. This 
is illustrated better in Fig.~2, where we show the relative differences 
$\Delta_{K_{m}}=L_{K_{m},19}/L_{19}$ with $L_{\alpha,\beta}=L_\alpha-L_\beta$ 
(the chosen $\sigma_R$-range starts at the $^4$He$(\gamma)$ break-up threshold). One again 
notes the very good convergence for $\sigma_R > 30$ MeV with almost identical 
results from $L_{13}$ to $L_{19}$. Altogether we consider our result for 
$\sigma_R > 30$ MeV as completely sufficient. On the other hand it is obvious 
that convergence is not entirely reached for lower $\sigma_R$. Here 
we should mention that for $L_{19}$ there are already 364000 states in the HH 
expansion. A further increase is beyond our present computational capabilities. 
On the other hand, as a closer inspection of Fig.~2 shows, the convergence   
proceeds with a rather regular pattern: 
(i) $L_{13,11} \simeq L_{11,9}$ and $L_{17,15} \simeq L_{15,13}$ and  
(ii) $L_{19,17} \simeq L_{17,15}/1.5 \simeq L_{13,11}/(1.5)^2$. 
Due to this regular pattern it is possible to obtain an 
extrapolated asymptotic result. We use the following Pad\'e approximation 
\begin{equation}
\label{Pade}
L^\infty_{K_{m}} = L_{K_{m-8}}+ L_{K_{m}-4,K_{m}-8}/
                         (1- \frac{ L_{K_{m},K_{m}-4} } 
                              { L_{K_{m}-4,K_{m}-8} } ) \,.
\end{equation}

In Fig.~3 we illustrate the results for $\sigma_{\gamma,K_m}$ obtained from the 
inversions of the corresponding transforms $L_{K_{m}}$. Due to the Lorentz kernel
the cross section presents the same features as $L$ itself: 
established peak height with a value very close to 3 mb, and not yet completely 
convergent peak position. In Fig.~3 we also show 
$\sigma^\infty_{\gamma,17}$ and $\sigma^\infty_{\gamma,19}$ resulting from the 
inversion of $L^\infty_{17}$ and $L^\infty_{19}$, respectively. 
Unquestionably, the extrapolated $L^\infty_{K_{m}}$ have a lower numerical quality than 
the calculated $L_{K_m}$ and consequently we do not find the same stability for 
the corresponding inversions (for details on inversion see \cite{ELO99}). 
Therefore we use an additional constraint in the inversion by fixing the peak 
cross section to the already converged value of 3 mb. 
In Fig.~3 it is evident that $\sigma^\infty_{\gamma,17}$ and $\sigma^\infty_{\gamma,19}$ 
are very similar, hence establishing a very good approximation for the asymptotic
$\sigma_\gamma$. One also notices that compared to $\sigma_{\gamma,19}$
they show a shift of the peak position by about 1 MeV towards lower energy.

In Fig.~4a we show our final results for $\sigma_\gamma$. Due to the 3NF one 
observes a reduction of the peak height by about 6\% and a shift of the peak 
position by about 1 MeV towards higher energy. Very large effects of the 3NF are 
found above 50 MeV with an enhancement of $\sigma_\gamma$ by e.g. 18, 25, and 35\% 
at $\omega=60$, 100, and 140 MeV, respectively (3NF effect could change somewhat 
if all multipole contributions are considered, see also 
discussion before Eq.~(1)). The comparison 
between the present results and those obtained with semi-realistic potential
models \cite{ELO97,BELO01} is similar as in the three-nucleon photodisintegration
\cite{ELOT}: semi-realistic models lead to rather realistic results in the giant 
resonance region overestimating the peak by about 10-15\% and giving 
quite a correct result for the the peak position; however, at higher $\omega$ the 
cross section is strongly underestimated (in the present case by a factor of three at 
pion threshold).

It is very interesting to compare the 3NF effects on $\sigma_\gamma$ to those 
found in the three-body nuclei \cite{ELOT,Fad_CHH}. Surprisingly, the 
reduction of the $^4$He peak height is smaller than in the three-nucleon case. 
For $^3$H/$^3$He its size is similar to the binding energy increase (10\%), 
whereas for $^4$He the 3NF increases the binding energy by 17\%, but reduces the peak 
by only 6\% and thus cannot 
be interpreted as a simple binding effect. Also at higher $\omega$ there are
important differences. The enhancement of the $^4$He cross section due to the 
3NF is significantly higher, namely about two times larger than for the 
three-body case. Interestingly this reflects the above mentioned different ratios between 
triplets and pairs in three- and four-body systems.
 
In Fig.~4b we compare our results to experimental data. For the AV18+UIX case  
upper/lower bounds are included to account for possible errors in the
extrapolation (\ref{Pade}). The bounds are obtained enhancing/reducing the 
difference $\sigma^\infty_{\gamma,19} - \sigma_{\gamma,19}$ by 50\%; we believe 
that this is a rather safe error estimate. 
The data in Fig.~4b have to be interpreted with some care: 
(i) in \cite{Wells} the peak cross section 
is determined from Compton scattering via dispersion relations, (ii) the dashed 
area corresponds to the sum of cross sections for $(\gamma,n)$ from \cite{Berman} and 
$(\gamma,p)^3$H from \cite{Feldman} as already shown in \cite{ELO97}, 
(iii) the data from the above mentioned recent $(\gamma,n)$ experiment 
\cite{Lund} are included only up to about the three-body break-up threshold, 
where one can rather safely assume that $\sigma_\gamma \simeq 2\sigma(\gamma,n)$ 
(see also \cite{Sofia1}), (iv) in \cite{Shima} 
all open channels are considered. One sees that the various data 
are quite different exhibiting maximal deviations of about a factor of 
two. The theoretical $\sigma_\gamma$ agrees quite well with the low-energy 
data of \cite{Berman,Feldman}. In the peak region, however, the situation is very 
unclear. There is a rather good agreement between the theoretical $\sigma_\gamma$ 
and the data of \cite{Lund} and \cite{Wells}, while those of \cite{Berman,Feldman} 
are noticeably lower. Very large 
discrepancies  are found in comparison to the 
recent data of Shima et al. \cite{Shima}. It is evident that the experimental 
situation is rather unsatisfactory and further improvement is urgently needed.

We summarize our work as follows. We have performed an ab initio calculation of
the $^4$He total photoabsorption cross section using a realistic nuclear force 
(AV18 NN potential plus the UIX-3NF). The full interaction is consistently 
taken into account
both in the ground state and in the continuum via the 
LIT method. For the solutions of the differential equations we use expansions in 
hyperspherical harmonics via the EIHH approach. Our results show a rather 
pronounced giant dipole peak. The 3NF reduces the peak height by only about 6\%, 
less than expected considering its large effect of almost 20\% on the $^4$He 
binding energy and its different role in the three-nucleon system. Beyond the giant 
dipole resonance 3NF effects become much larger. With growing $\omega$ their 
importance increases and at pion threshold one has an 
enhancement of 35\%, about twice the effect one finds in $^3$H/$^3$He 
photoabsorption. Close to threshold the theoretical cross section agrees quite 
well with experimental data. In the giant resonance region, where there is no 
established experimental cross section, our results are in  good agreement 
with the data of \cite{Lund} and \cite{Wells}, while we find a strong 
disagreement with the data of \cite{Shima}.

In conclusion we would like to emphasize that it is very important to 
understand whether a nuclear force model, which is constructed in the two- and 
three-nucleon systems, is sufficient to explain the four-nucleon
photodisintegration. To this end further experimental investigations are
mandatory.

This work was supported by the Israel Science Foundation (grant no 202/02) 
and by the Italian Ministry of Research (COFIN03). Numerical calculations were 
partly performed at CINECA (Bologna); one of us (S.B.) 
thanks S. Boschi (CINECA) for his professional help.

\vfill\eject

\begin{figure}
\resizebox*{15cm}{17cm}{\includegraphics[angle=-90]{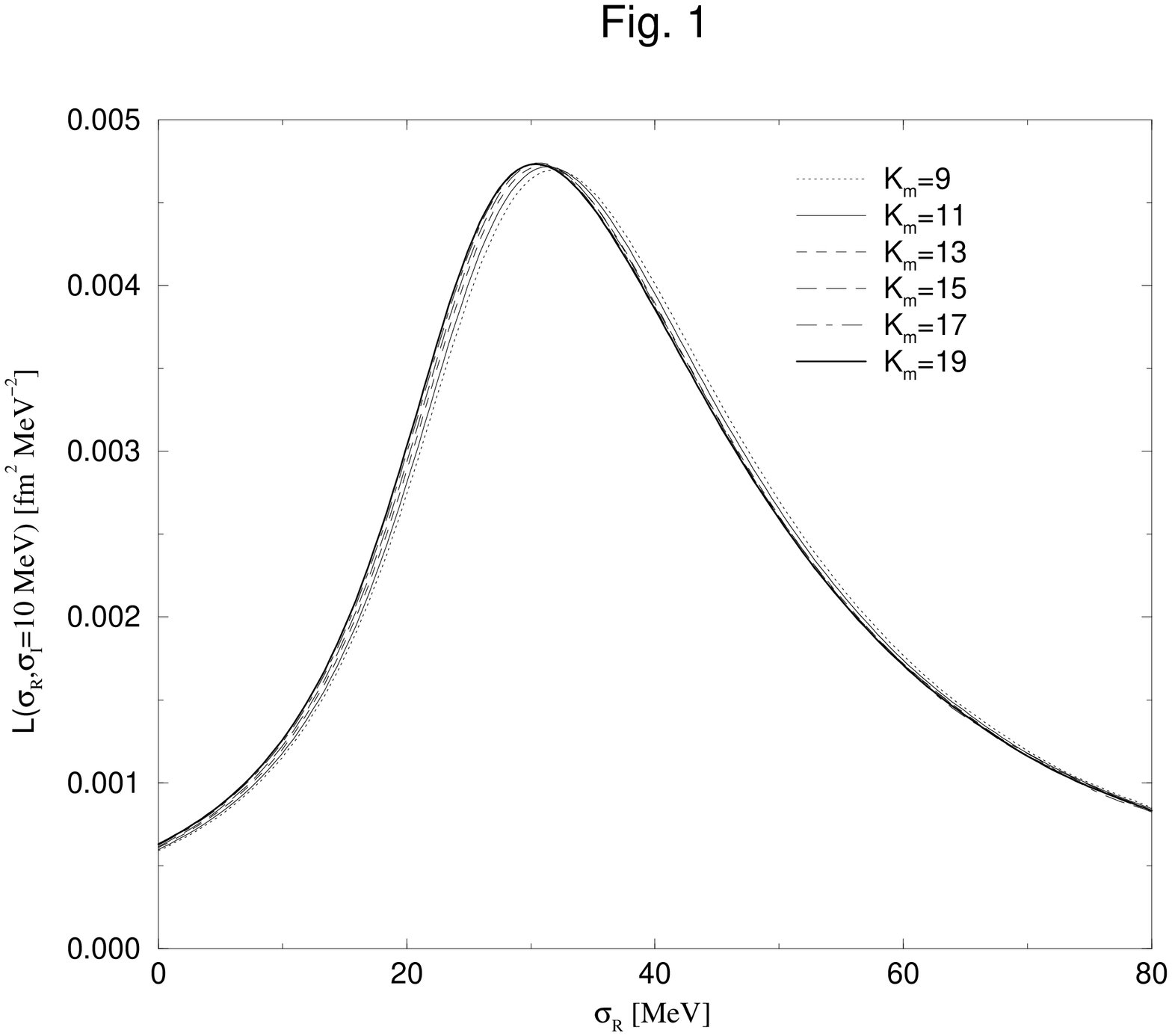}}
\caption{Convergence of $L_{K_{m}}$ with $\sigma_I=10$ MeV (AV18+UIX).}
\end{figure}

\begin{figure}
\resizebox*{15cm}{17cm}{\includegraphics[angle=0]{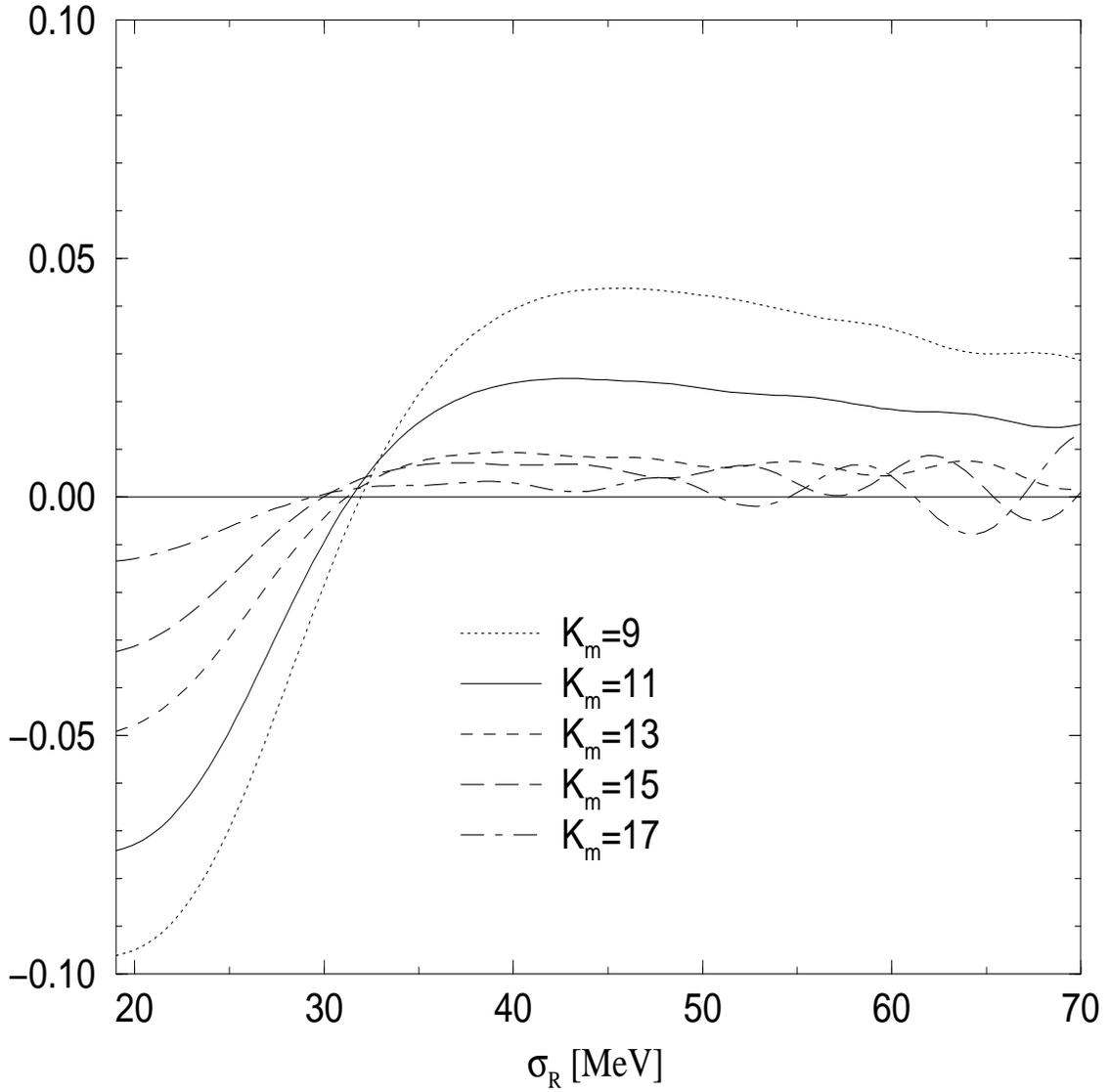}}
\caption{Convergence of $\Delta_{K_{m}}=(L_{K_m}-L_{19})/L_{19}$  (AV18+UIX).}
\end{figure}

\begin{figure}
\resizebox*{15cm}{17cm}{\includegraphics[angle=-90]{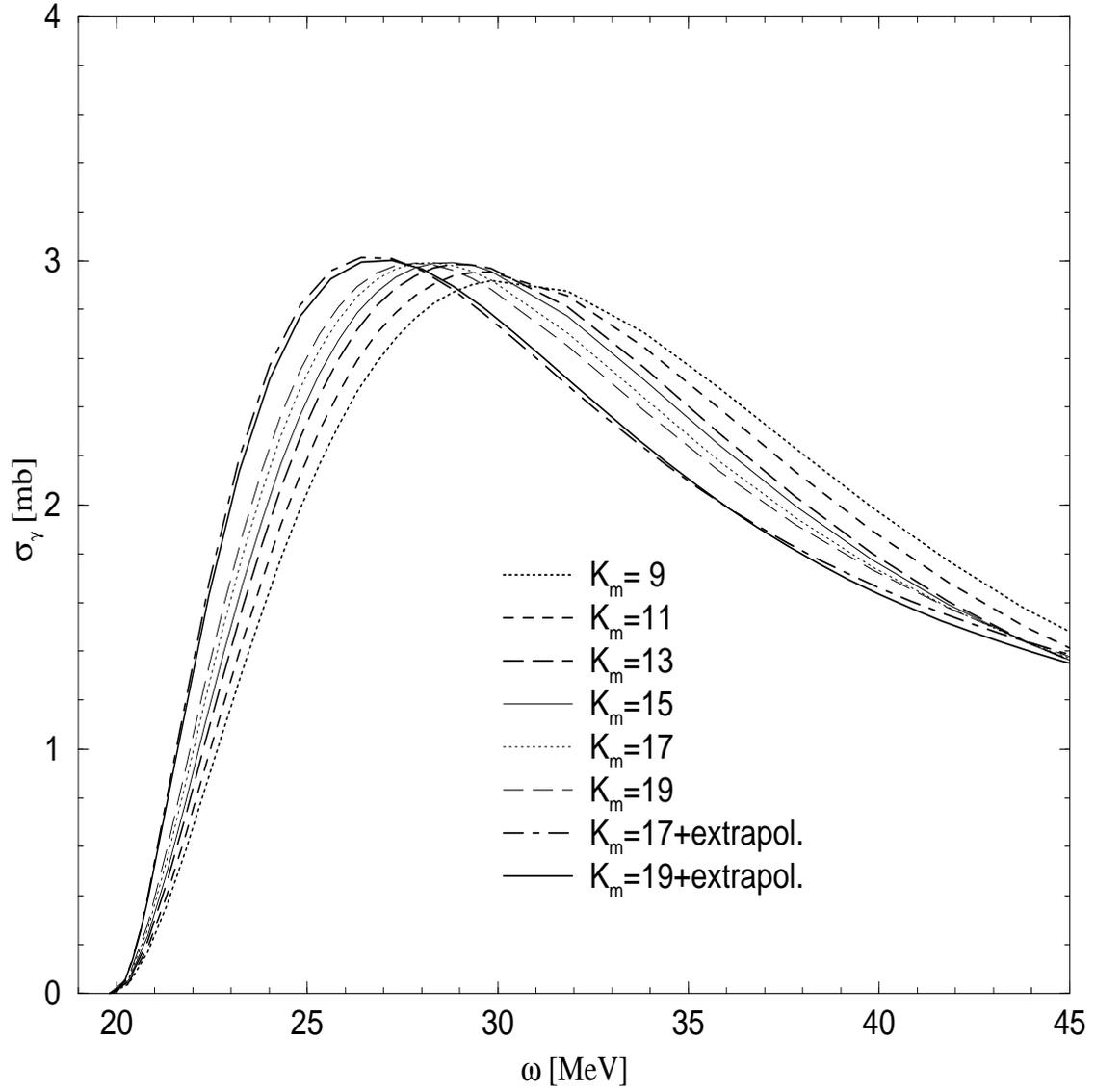}}
\caption{Convergence of $\sigma_{\gamma,K_m}$ (AV18+UIX), also
shown  $\sigma^\infty_{\gamma,17}$ and $\sigma^\infty_{\gamma,19}$.}
\end{figure}

\begin{figure}
\resizebox*{15cm}{17cm}{\includegraphics[angle=0]{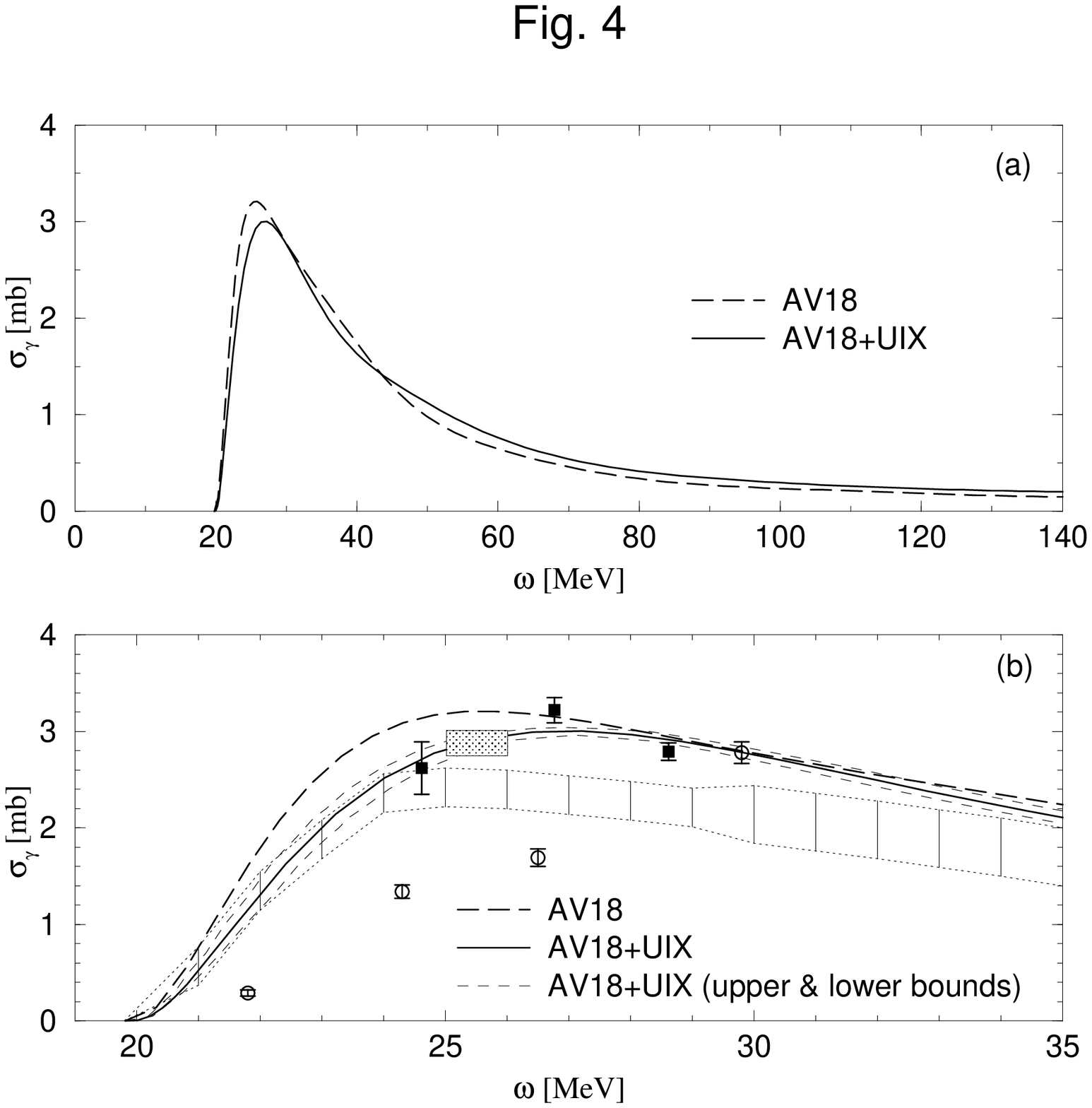}}
\caption{Total $^4$He photoabsorption cross section: (a) $\sigma_\gamma$ (AV18) 
and $\sigma^\infty_{\gamma,19}$ (AV18+UIX),
(b) as (a) but also included upper/lower bounds and various experimental 
data (see text), area between
dotted lines \cite{Berman,Feldman}, dotted box \cite{Wells}, squares 
\cite{Lund}, and circles \cite{Shima} .}
\end{figure}

\begin{table}
\caption{
\label{tb:4he_gs}
Convergence of HH expansion for $^4$He binding energy $E_b$ [MeV]  
and root mean square matter radius $\langle r^2\rangle^{\frac{1}{2}}$ [fm] with AV18 and 
AV18+UIX potentials. 
}
\begin{center}
\begin{tabular}
{c | cc | cc} 
               & \,\,\,\,\,\,\,\,\,\,\,\,\,\,\,\,\,\,\, AV18 &  
               & \,\,\,\,\,\,\,\,\,\,\,\,\,\,\,\, AV18+UIX & \\ 
    $K^0_{m}$ & $E_b$ & $ \langle r^2\rangle^{\frac{1}{2}} $ 
              & $E_b$ & $ \langle r^2\rangle^{\frac{1}{2}} $ 
              \\ \hline \hline
  6 &  25.312  &  1.506  & 26.23   &  1.456  \\ 
  8 &  25.000  &  1.509  & 27.63   &  1.428  \\
 10 &  24.443  &  1.520  & 27.861  &  1.428  \\
 12 &  24.492  &  1.518  & 28.261  &  1.427  \\
 14 &  24.350  &  1.518  & 28.324  &  1.428  \\
 16 &  24.315  &  1.518  & 28.397  &  1.430  \\
 18 &  24.273  &  1.518  & 28.396  &  1.431  \\
 20 &  24.268  &  1.518  & 28.418  &  1.432  \\ \hline
\end{tabular}
\end{center}
\end{table}
\end{document}